# ScrollyVis: Interactive visual authoring of guided dynamic narratives for scientific scrollytelling

Eric Mörth, Stefan Bruckner, and Noeska N. Smit

**Abstract**—Visual stories are an effective and powerful tool to convey specific information to a diverse public. Scrollytelling is a recent visual storytelling technique extensively used on the web, where content appears or changes as users scroll up or down a page. By employing the familiar gesture of scrolling as its primary interaction mechanism, it provides users with a sense of control, exploration and discoverability while still offering a simple and intuitive interface. In this paper, we present a novel approach for authoring, editing, and presenting data-driven scientific narratives using scrollytelling. Our method flexibly integrates common sources such as images, text, and video, but also supports more specialized visualization techniques such as interactive maps as well as scalar field and mesh data visualizations. We show that scrolling navigation can be used to traverse dynamic narratives and demonstrate how it can be combined with interactive parameter exploration. The resulting system consists of an extensible web-based authoring tool capable of exporting stand-alone stories that can be hosted on any web server. We demonstrate the power and utility of our approach with case studies from several of diverse scientific fields and with a user study including 12 participants of diverse professional backgrounds. Furthermore, an expert in creating interactive articles assessed the usefulness of our approach and the quality of the created stories.

## 1 INTRODUCTION

STORYTELLING is deeply embedded in our society. Its purpose ranges from informing and recording to explaining and entertaining. Stories play a major role in understanding the world, cultural identity and to trigger and explain emotions. Stories can be told in many different ways including passive and interactive forms as well as linear and nonlinear methods [20]. Filmmakers explore a vast range of methods to convey a story in the best possible way. Quentin Tarantino is an example of a film director who is known for exploring new ways to present a story and therefore his movies are frequently discussed [20]. Stories are an important aspect not only of movies but also in the context of video games, books, and articles. Newspaper article authors engage the viewer by including graphics and images but on the web they can even add interaction [36].

With the rise of the internet, storytelling became a part of conveying information in online media. In more recent times, storytelling on the web was re-imagined into so-called scrollytelling [37]. Viewers interact with presented information not by clicking through a web slideshow but by simply scrolling through a website. The author can plan the flow of information and is able to guide the reader through the experience. Scrollytelling is extensively used by news outlets such as the New York Times [33], [35] as it can be an engaging and effective way to present information. On mobile devices in particular, scrollytelling is one of the standard interactions [33] and heavily used by social media platforms and various online media.

Storytelling can be a powerful tool to convey scientific information [23]. Science has to compete with other storytellers, many of whom are not bound to scientific evidence [9]. Blastland et al. [2] referenced the philosopher Onora O'Neill who said: "inform but not persuade" and when sharing evidence, suggested to strive "to be accessible, comprehensible, usable and assessable".

Scientific communication must preserve its credibility, but also needs to engage the audience with compelling communication formats. Dahlstrom [8] noted that storytelling often has a bad reputation in science and that there is even a mantra saying "The plural of anecdote is not data". He proposed the following adaption to this statement: "The plural of anecdote is engaging science communication". Scientific storytelling can leverage data visualization to convey outcomes and facilitate reasoning about scientific results, serving goals such as the communication with peers in the field or engaging a wider community.

Hohman et al. [18] demonstrated that interactive articles can be used to present the latest progress in various research fields and to make the findings accessible and understandable to a broad audience. The challenge in doing so is that often there is no clear incentive structure as well as little funding for research dissemination and communication. Interactive and engaging articles are most viable on the web [18]. One challenge is that not all scientists possess web-development skills. There is a variety of editors and content management systems to create basic websites, but interactive and engaging storytelling is not supported sufficiently in these. As demonstrated by the user evaluation of Seyser et al. [37], a whole team including authors, designers, and developers is needed to create such rich experiences. In terms of dynamic narratives, authoring tools facilitating storyboarding are currently available, but scrollytelling support is lacking. At present, there is a missing link between dynamic narratives and scrollytelling presentation on the web.

Our approach aims to fill this gap. With this paper, we present ScrollyVis, an extensible web-based authoring tool for creating guided dynamic narratives with a particular focus on scientific narratives. ScrollyVis offers a processing pipeline which exports authored stories such that they are ready for deployment on any web server. We support dynamic as well as static narratives. Our approach enables users of all technical skill levels to create scrollytelling web experiences with ease. We allow for the integration of a broad range of visual media, such as images, videos, and map views, but also accommodate more advanced

• Eric Mörth, Stefan Bruckner, and Noeska Smit are with the Dept. of Informatics, University of Bergen and the Mohn Medical Imaging and Visualization Centre, Dept. of Radiology, Haukeland Univ. Hospital, Norway. E-mail: eric.moerth@uib.no, stefan.bruckner@uib.no, noeska.smit@uib.no

visualization techniques such as direct volume rendering, slice-based visualization and 3D surface-based visualizations. To verify the utility of our approach, we present a user evaluation including 12 participants with diverse professional backgrounds. Furthermore, we present four case studies in collaboration with experts from a range of scientific disciplines and a qualitative evaluation of our approach and the resulting stories by an expert in interactive storytelling on the web. Our main contributions in this paper are:

- We introduce a system which allows for efficient authoring, generation, and presentation of dynamic media-rich scrollytelling experiences on the web for scientific communication.
- We present ScrollyVis, a prototype storyboard-based editor realizing the system description enabling users to author and publish dynamic narratives on the web without requiring prior web development skills.
- We demonstrate the power and utility of the approach through case studies and evaluate the usability of the editor and potential through user feedback.

## 2 RELATED WORK

In this section we describe work related to our approach, such as storytelling in visualization and visual exploration. Furthermore, we reflect on scrollytelling as a web-based long-form article, and storytelling editors.

**Storytelling in Visualization**: Storytelling is a focus in visualization research over a longer period of time. Wohlfart et al. [50] combined storytelling with interactive volume visualization to enable a better understanding of the underlying data and information. They introduced an authoring and a storytelling step to separate exploration and presentation of a story. Their approach employed Shneiderman's [38] information seeking mantra and proposed a taxonomy for interaction with the user. Furthermore, they presented a story model consisting of story nodes and story transitions. Building up on their efforts, we introduce story segments as a higher-level abstraction of semantically similar story nodes.

Kosara et al. [23] argued that stories are a good way to present data as they package important information and knowledge in an easily understandable way. They highlight that interaction is one of the most important aspects of visualization, including altering the pace and direction of the story. Ma et al. [31] reflected on the question of what "good pacing" means when it comes to scientific storytelling using visualizations. Every spectator has their own pacing preference and there is always a compromise when introducing a fixed pace to tell a story. The authors furthermore emphasized the importance of user domain knowledge and relevance of the story to the users. The paper is especially relevant for our approach as it guides our design in pacing and adapting stories to a target audience. Tong et al. [45] describe common visualization types used in storytelling. Based on their analysis, we designed our system to support all of these visualization types in order to make them readily available for scientific communication. In the following, we present visualization approaches which specifically aim to tackle challenges related to scrollytelling.

**Scrollytelling**: According to Pimbaud [35], scrollytelling has been around since 2010 and is linked to the success of social media. He further speculates that scrolling might be the easiest user interaction possible. Seyser et al. [37] claim that scrollytelling is the web equivalent to long-form articles used in journalism. Scrollytelling presentations frequently consist of multimedia content and information visualization in particular. The authors stated that the narrative structure of scrollytelling articles is either linear or elastic. The latter enables the user to dive deeper into the story on demand. Scrollytelling articles often use at least three different multimedia elements including photography, videos, and visualizations [17], [47]. Pettersen [33] proposed that storytelling is not only about the words but presenting information in the most interactive and exciting way. Furthermore, she mentions that scrollytelling is beneficial to engage and actively keep the viewer's attention during story consumption. Scrollytelling gives a sense of control, exploration, and discoverability.

Unfortunately, to the best of our knowledge, there are no publications on concrete guidelines for best practices in scrollytelling. However, practical recommendations are available from various blogs. For example, McKinley [34], principal engineer at Etsy, shared that endless scrolling is unfavorable and resulted in lower sales at Etsy. Furthermore, Seyser et al. concluded that the Gestalt laws should be applied to structure visual elements in order to make them understandable to the viewers. Kosara [22] provides an extensive overview of what he considers bad scrollytelling design features and demonstrates that many aspects can easily annoy and distract viewers of scrollytelling websites. Kosara mentions that the user should know upfront how long the story is going to be and that direct access to different parts of the story is favorable [22]. Scrollytelling content needs to be carefully designed. Transforming existing websites by simply adding scrolling and unveiling content on demand often ends up in an unsatisfactory user experience. Bostock [3] proposed scrolling implementation guidelines, which equip us with five rules to follow in order to create an effective scrollytelling website, namely: *(1) Prefer scrolling to clicking, (2) allow rapid, incremental, reversible scrolling, (3) provide instantaneous consistent feedback, (4) avoid unwanted disruptions and (5) support standard keyboard controls.*

**Narrative Authoring and Storytelling Editors**: According to Hohman et al. [18], creating interactive articles today is still difficult. It is often closer to building a website than to writing a blog post or article. It also takes considerably more time than writing a static article or even a scientific publication. According to Conlen et al. [6], media such as the New York Times, Washington Post, the Guardian, and FiveThirtyEight, provide high quality multimedia narratives often referred to as interactives. The authors noted that the data visualization community suggested research opportunities in creating tools for driving the production of such interactive narratives. Creating or re-creating interactives as presented by high quality newspaper providers is complex and involves several experts [6].

Tableau [15] implemented their story feature by using story points. Story authors can present interactive visualizations created in Tableau in a slideshow manner. While the visualizations within one story point are interactive and allow for exploration, the transition between points is not dynamic. In comparison to their approach, we focus on smoothly animated transitions that change visual elemements in an incremental and reversible manner. Furthermore, we allow for 3D data visualization, such as volume and surface visualizations. Kouřil et al. [24] presented a novel way to prepare story structures and automatically create concrete narratives for molecular documentaries. They present a technique called *story graph foraging* and techniques for real-time narrative synthesis. In contrast to their approach, we provide our stories on the web and not as video and have a broader focus than molecular visualization. *VizFlow* by Sultanum et al. [43] instead focused on data-driven articles. The authors used text-chart linking strategies to create scrollytelling experiences and evaluated their approach with 12 authoring and 24 reading participants. Compared to *VizFlow*, we do not focus on data visualization exclusively and allow for more extensive authoring opportunities. In addition to narrative videos and text enriched with data visualization, interactive data comics are another approach for presenting scientific insights. Wang et al. [49] presented a lightweight specification language entitled *Comic Script* to create interactive and dynamic data comics. The approach supports branching, change of perspective, and details-on-demand, as interaction methods for the viewer. The approach has similar dynamic capabilities as our approach, but leverages a slideshow-like format, employing point and click user interactions. Furthermore, we enable the visualization of more complex data types such as maps, surfaces and volumes.

Two well known software applications for creating dynamic stories are *Stornaway* [42] and *Twine* [46]. *Stornaway* features a node-link editor which allows for the creation of dynamic interactive videos. *Twine* is an open-source tool for telling interactive, nonlinear stories. *Twine*'s editor is based on a node-link diagram and is designed for the creation of interactive fiction. *Twine* exports the story directly to HTML. In comparison to these two applications, our approach enables the creation of stories including various additional media types and creates a scrollytelling website rather than a click-based website or video-based presentation, as is the case with *Twine* or *Stornaway*, respectively. Furthermore, our editor includes a preview of all different media content and introduces a hybrid approach between node-link diagrams and storyboards. In contrast to these approaches based on graphical user interfaces, Satyanarayan et al. [36] introduced a system which combines a domain specific language (DSL), Ellipsis, with a graphical user interface-based story authoring tool. The authors contributed a model for narrative visualization which helps story authors who may not be familiar with web development to convey their stories as websites. They evaluated their approach with a qualitative user study with feedback from journalists. The journalists where positive overall but mentioned that a node-link interface would be a good way to author a story. Furthermore, the journalists asked for an easier way to present non-linear stories. In contrast to this related work, we use scrollytelling to guide the user through the story. Furthermore, we provide a node-link authoring tool and support guided dynamic narratives.

The closest previous work to our approach is *Idyll* [6], [7] which consists of a markup language for authoring and publishing interactive articles on the web. This approach is based on a DSL designed for authoring interactive narratives combining a markup language and in-line JavaScript components. *Idyll Studio* [7] has a graphical user interface that lowers the threshold for non-experts to create interactive articles. Compared to *Idyll*, we do not focus on creating a DSL and rather concentrate on editor functionality and flexible support for different visualizations methods. Furthermore, in addition to using parameter-based interaction we allow for dynamic narratives where the viewers can choose pre-defined narrative paths.

## 3 SCROLLYTELLING AND NARRATIVES

Storytelling is an effective way of conveying information and knowledge [28]. According to Joubert et al. [19], storytelling is the soul of science communication. Storytelling and especially scrollytelling is increasingly used on the web. This trend is also reflected in news outlets such as The Economist, the BBC, the New York Times and German science magazine Substanz. Two examples of such stories are Unearthing the Truth [13] and Genexpressionen [29]. In addition to stories created by news agencies, Apple research recently started to present their latest research in form of interactive scrollytelling based web pages, e.g., a story about interpretable adaptive optimization [41]. All of these stories are a joint effort between authors, designers and programmers to enable media-rich and interactive scrollytelling experiences.

Across these sample stories, we have identified several common patterns. All of the stories include textual information, in most cases combined with media such as images, videos and audio. In some cases, more complex visualizations are presented, such as the surface visualization of photogrammetry data in the story by the Economist [13]. The stories can be further categorized into partial or full scrollytelling websites. One example of a full scrollytelling website is the the Genexpression story [29], while the Economist story [13] or the story created by Apple Research [41] employ it partially. Creating such stories is associated with considerable costs and frequently involves a team of programmers, web developers, authors, designers, directors and content consultants, as exemplified by the the Genexpression story [29]. One of our main goals is to reduce this cost and efforts in order to enable much smaller teams or even single individuals to create immersive and impactful stories about their work or other topics of interest.

Stories can be told in many ways, influencing how the story is perceived by the listeners, viewers or readers. A narrative specifies the order in which events are told as opposed to the order the events actually happened [16]. Narratives can be in chronological order, telling the story in the order it happened, but it is also possible to present events out of order in a nonlinear way [20]. In contrast to the story itself, which focuses on the content, narrative is the expression of a story [44]. Narratives can be used to maintain a sense of mystery by, e.g., withholding information in order to keep tension high and the audience engaged [20]. User engagement is a very important aspect in storytelling and to this end user-directed paths can be employed [45]. We differentiate in this way between passive and interactive stories. To prevent ambiguity by overloading the term "interactive" in a visualization context, we will refer to such stories as "dynamic" in our approach.

Authoring media-rich scrollytelling experiences mainly consists of two parts: defining the content to present and specifying the relationships between content items. This can be achieved in many different ways, as existing editors like Stornaway [42] and Twine [46] demonstrate. Based on the discussions in the literature as well as our analysis of scrollytelling content, we identify the following requirements for a system enabling authors to create such stories:

1) The system should enable efficient definition of story content and content item relations.
2) To support multiple narrative structures, there needs to be support for story branching.
3) The system should ideally support all narrative structures proposed by Munday [32].
4) The system should enable authors to specify how content items representing different or same media types should be combined.

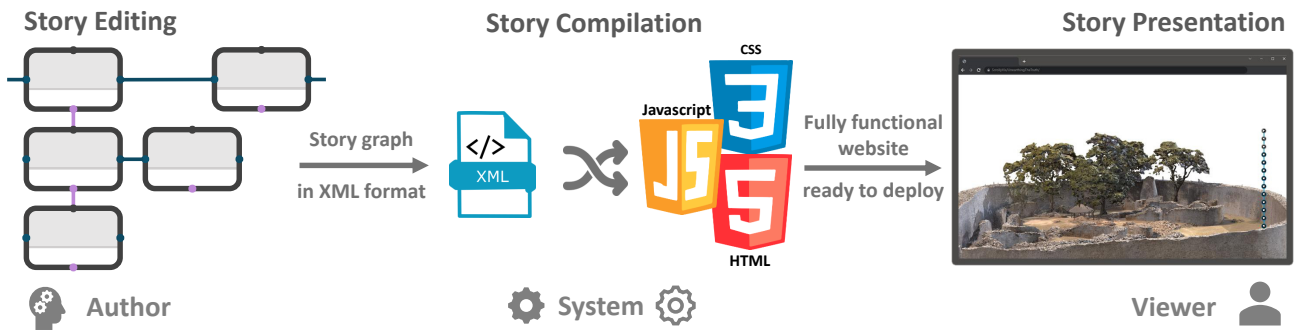

Fig. 1. The ScrollyVis system consists of three main parts: story editing, story compilation and story Presentation. The editing phase allows authors to create complex dynamic narratives in the ScrollyVis editor. The story is exported in an XML format and is compiled into a fully functional website within the ScrollyVis compilation phase. The website can subsequently be presented in any standard web browser while the ScrollyVis created code handles the dynamic narrative handling.

5) Complex content layering combinations should be possible in the system.

6) The system should be easy to use for users at various skill levels and allow for efficient story creation.

7) The created stories should be viewable in a standard web browser.

8) The story should support rapid incremental and reversible scrolling interaction.

## 4 ScrollyVis

To support the creation of media rich scrollytelling experiences, we introduce ScrollyVis, a web based story authoring system which follows a no-code paradigm to create media-rich and dynamic narratives. In the following, we present the individual concepts and components of the authoring system. The system has three main components responsible for story editing, compilation, and presentation (see Figure 1). In the following, we outline the main concepts in these three components.

### 4.1 Story Editing

The first component of our system is the story editor. This part of the system is responsible for the specification of story content and the definition of relations between the content items.

Wohlfart and Hauser [50] employed the concepts of story nodes and story transitions in their visual storytelling approach. Story nodes are stages within a story where content is presented. They are connected by transitions and, similar to the work by Wohlfart et al. [50], these transitions ensure a gradual progression between story nodes and fulfill requirement 8 from Section 3. Instead of clicking through a slide show, we use scrolling as the only interaction control of the story progress which follows rule 1 of Bostock [3]. We abstract story nodes to consist of three different parts as visualized in Figure 2 A: a node preview, node parameters, and node connections. In our system a story node is used as a content item and fulfills requirement 1 as outlined in Section 3. This design was motivated by storyboards which are widely used for the purpose of pre-visualizing a movie, animation, or interactive media sequence. Depending on the node type, the preview can be static or interactive. In the current version of ScrollyVis, nodes can consist of the following media types: text, image, video, audio, map, 3D volume visualization and 3D surface visualization. Nodes which allow for camera control such as volume or surface visualizations provide corresponding interaction facilities within the preview window. Static previews are used for images, videos, text, and audio files. The second part within a story node are the parameters. The node parameter set is dependent on the type of visualization. While images, for instance, have parameters specifying the position within the website and the size of the presentation, complex nodes like volume visualizations feature a larger set of parameters. These include the definition of which volume visualization method shall be used, e.g., ISO value, maximum intensity projection or direct volume rendering with a transfer function, and associated parameters. To support rapid story authoring, we initialize every story node with a pre-defined set of sensible default parameters which can be altered as needed. Story nodes feature connections ports, which are used to control the story flow. In our system, every node is associated with a pre-defined code segment which is instantiated with the given parameters. This instance is used when the story author places the node onto the canvas of the editor. This allows for easy extension as new node types simply have to be defined as a new template code in order to be usable in the editor. Story nodes can be connected from left to right and from top to bottom. Every node therefore features an output port on the right and on the bottom side and an input port on the left and top.

**Layers:** To create complex combinations of story nodes we allow authors to combine different nodes in a layered manner. Authors have to be able to specify that one story node is presented while another story node is still shown to support media layering. There are different ways to depict such a behavior in an editor. Inspired by layered tracks used in common video editing software, we propose a sub-path feature as visible in Figure 2 B. This sub-path allows authors to link story-nodes not only going from left to right but also from bottom to top. Story nodes connected at the bottom of a previous node are added as another layer in the final story result. Story nodes attached on the right side of previous node replace the previous one. With this simple design, different media types can be combined easily. In the example in Figure 2 B, the story elements are presented as follows. Item A is presented first, and item B is presented while A stays visible. C is presented while A and B stay visible. D is presented while A stays visible, and B and C disappear. Finally, E gets shown as a new content item. This fulfils requirements 4 and 5 discussed in Section 3. We refer to a node together with its sub-path as a *story segment*. Story segments encapsulate parts of the story that belong together and are used as the primary navigational unit in the presentation of a story.

**Dynamic Narratives:** Scrollytelling on the web is characterized by incrementally revealing information based on scrolling interaction by the viewer. In scrollytelling, viewers traverse a webpage along one axis in a linear fashion. The interaction is constrained to a single degree of freedom, i.e., scrolling up or down. When introducing dynamic narratives this pattern is no longer sufficient. Dynamic narratives are a powerful story telling tool which boosts viewer engagement as story immersion increases [32]. To support dynamic narratives and therefore requirement 2, we introduce decision nodes and allow for multiple story endpoints in our node-link storyboard as depicted in Figure 2 C. At decision nodes, the author determines a point in the story where the story branches out and the viewer is able to decide the next story segment. The author defines several story path options and defines the available branches. By introducing such decision nodes we also fulfill requirement 3 introduced in Section 3.

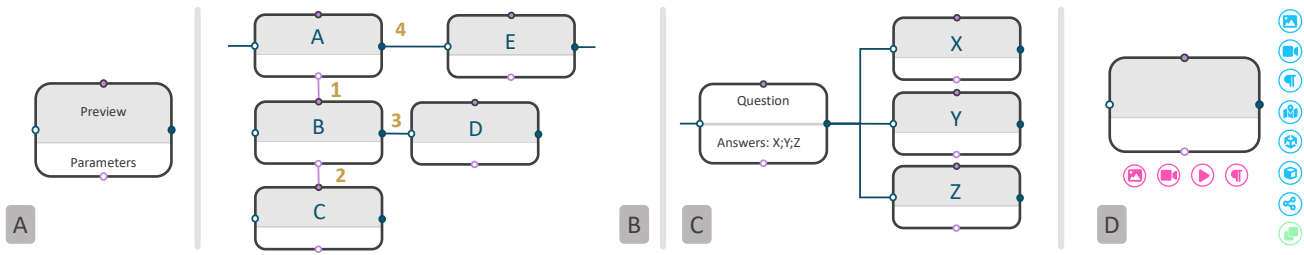

Fig. 2. A: presents the abstract unit story node which consists of a preview, a properties section and connection nodes on all four sides. In B story transitions are depicted, in this example the layering possibilities in ScrollyVis are shown. C presents a decision node where the viewer can later in the story presentation choose which path they would like to pursue and in D the context menu is depicted which allows for fast and accurate story creation.

**Interface:** Our prototype web-based editor was designed with efficiency and ease of use in mind, according to requirement 6 from Section 3. We allow users to build a scrollytelling website by visual programming, i.e., dragging and dropping story nodes onto a storyboard editor or by using a context menu shown and establishing links. We introduce several usability features to allow for efficient story editing. First, every connection port on a story node features a context menu which shows the possible new node type connections on demand. The context menu is presented in Figure 2 D. Via the bottom connection menu, an image viewer, video player, audio player or a text node can be created. When using the context menu on the main path, every node type is available. As soon as the user selects one of the possible node types, the new node is automatically placed on the canvas at a fixed offset to the current node and on the same horizontal or vertical position, depending on whether it is a sub-path or main path connection. Furthermore, editor canvas automatically scrolls to focus on the new node. This allows for fast and accurate story creation with minimal interaction, while the story graph layout is optimized by default.

In the context menu along the main path, we also provide a copy functionality as present in Figure 2 D in green. This feature allows the story author to copy the current node with all its properties. The copy is automatically connected to the current node. This feature is mainly interesting for node types which feature complex interaction methods like camera control (e.g., volume or surface visualizations). By copying the current node with all its settings, animated camera transitions can be authored in a rapid manner. In the copied node, the author just has to move adjust the settings and in the final web page the camera will zoom, pan, and rotate to transition between the two views.

The story graph is serialized into an XML format. The XML document stores all nodes and their parameters as well as all node connections. The resulting files are human readable and can also be adjusted manually. Furthermore, this simple intermediate format also enables the future exploration of additional higher-level interfaces such as wizards which may serve as a starting point for customization. After the author finishes story editing, they can export the story as a web page ready to deploy it on a web server. We call this process story compilation and describe it in the following subsection.

## 4.2 Story Compilation

At any point during the editing phase, the user can trigger the compilation of the story into a ready-to-deploy website. The input of the story compiler is the story graph in XML format together with all media content, e.g., images and videos. Depending on whether decision nodes are present, a story can be presented as a linear sequence of story nodes or as a tree where the narrative structure branches out after every decision node. Every story node represents different types of content, e.g., text or an image, that has to be translated from story logic in XML to HTML, CSS and JavaScript code for the final presentation in every standard web browser which fulfills requirement 7 presented in Section 3. The compilation phase is divided into two steps: first the story content nodes are created and then the node transitions are set up.

**Story Contents:** First, the story graph delivered from the ScrollyVis editor is traversed to define the story tree and to create the content for the resulting website. Each story node is represented as pre-defined HTML and CSS code, dependent on the node media type. Less complex media like text, images, videos, and audio can be converted directly to HTML code. More complex media, e.g., map views, volume visualization, and surface visualization, need content loading code in addition. On traversal of the story graph, a story tree is created. Except for the root, leaf, and decision nodes, every node has exactly one predecessor node and one successor node. In this step the sub-tree of every node connected via the sub-path port is flattened. The sub-path feature is especially important in defining the transition between story nodes. A decision node is the only node which has multiple successor nodes to create branching points in the story tree. In a second step, we traverse the generated story tree and take main- and sub-path information into account.

**Story Transitions:** In the second step of the story compilation, we focus on story node transitions. The behavior of every node in the story is defined by its state before, during, and after node traversal. Before traversal, the node content is not visible. This state is defined in the JavaScript code by setting the opacity to 0. Node transition handling depends on the existence of a sub-path. If there is no sub-path, the node will be blended in and out within the scrolling extent of the current node. In our current version we a scroll extent of 1000 pixels per story node by default, but this value can be adjusted by the author. If there is a sub-path, the current node will be blended in and then faded out only when the next main path node is traversed. In addition to blending node opacity, further transitions such as camera movement and dynamic parameter changes are also handled within the node scrolling extent. After the node has been traversed, it indicates its successor node. The only exception to this is the decision node. At a decision node, successor selection is dependent on viewer interaction. As soon as the viewer decides upon a story branch and scrolls, the selected successor node is hooked to the decision node and the story viewer can carry on viewing the story. When scrolling back up the successor is unhooked again to enable different story path traversals. The compiled story includes node content in HTML and CSS as well as node content loading and transition code in JavaScript. In addition, it includes static code handling volume visualization, the Sketchfab integration for surface rendering, Mapbox code for map views, and static CSS styling code. The compiled story is put together as a complete website that is ready to deploy on a web server.

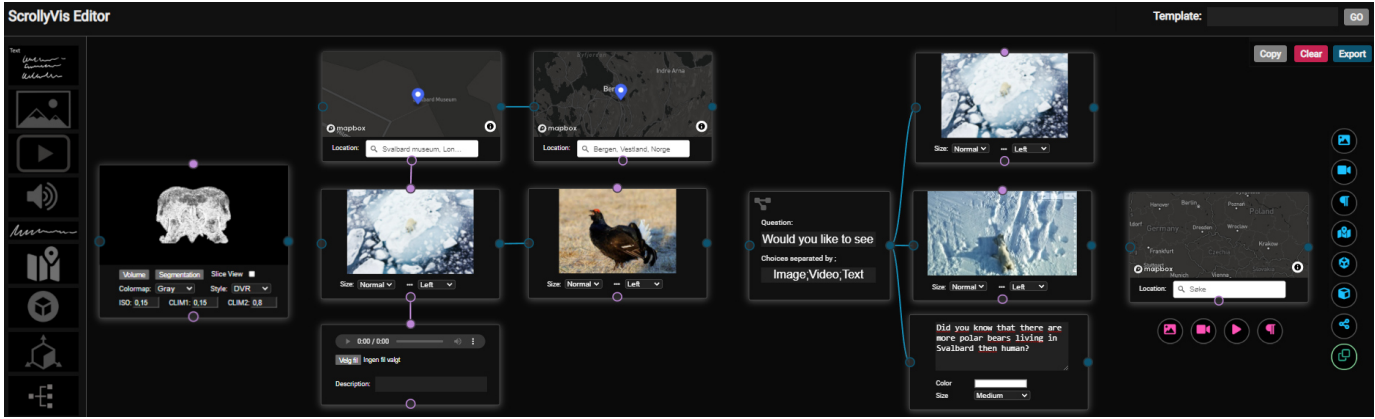

Fig. 3. Depiction of the ScrollyVis editor representing the same functionalities as presented in the abstract system depiction in Figure 2. Stories can be created either by dragging nodes from the left onto the canvas or by using the context menu as presented on the right side of the Figure. All nodes where applicable feature a preview window.

### 4.3 Story Presentation

After compilation, the story is ready for presentation. Stories created in the ScrollyVis editor are exported ready to be deployed on a webserver after compilation. Viewers can simply open the scrollytelling story in a web browser of their choice and interact with the story by scrolling. One important aspect of smooth story presentation are the transitions between different story nodes. In our case, the transition between the nodes is defined automatically, based on the current node content and content of the story segment displayed previously. These transitions allow for rapid, incremental, reversible scrolling and to provide consistent and instantaneous feedback following rule 2 and rule 3 from Bostock [3]. In addition, we introduced a rudimentary keyboard control option to progress the story to fulfill the last rule of Bostock [3] which could be extended on in a future version of ScrollyVis.

### Transitions

Node transitions depend on the predecessor and successor node content and parameters. Furthermore, behavior differs according to whether they are part of main- or sub-path traversal. In general, as sub-paths are traversed, transitions are made by blending the opacity of the current and previous node content. Within the main path more elaborate transition methods may be used. Between media nodes of a different type, e.g., picture to video or volume to surface visualization, the transition works via opacity change. More elaborate transition methods are available for map, volume, slice, and surface nodes. Every transition is linked to viewer scroll interaction and works in both scroll directions without using triggered animations. One exception is the embedding of videos, where video playback is triggered by viewer scrolling interaction.

**Map View**: When first entering a map view from any other node type, the map will be introduced by an opacity change combined with a zooming transition from far away to close to the target location. Between two map views at different locations, the story viewer will fly from the initial location to the next target location. On this flight, the zoom level is adjusted to replicate a parabolic flight or jump [48].

**Direct Volume Rendering and Slice Views**: When entering a volume visualization node, the zoom level is adjusted to the zoom level set by the story author in the volume visualization node. When transitioning from one volume visualization node to another with the same input data, multiple parameters are linearly blended. Camera translation, rotation, and zoom level are updated based on user scrolling interaction. In addition, there are volume-specific parameters available for blending such as the lower and upper intensity value range limit and the ISO value. The lower and upper intensity value range limits are used to change the contrast for the volume rendering by limiting the applied color map to a reduced value range. The ISO value defines the value which is used in the ISO-Surface volume rendering method and defines which voxels are set to be within and which ones are outside of the surface of interest. If the volume is shown as a slice view, the slice index is also altered from the previous node slice index to the current one.

**3D Surface Visualization**: When entering a 3D surface visualization node, the opacity of the visualization is altered based on viewer scrolling interaction. Between two 3D surface visualization nodes representing the same data set, similarly to volume rendering, camera position, camera rotation, and camera zoom level are updated through scrolling. These transitions allow for natural camera movements and animations by simply defining start and end camera positions in two story nodes within the ScrollyVis editor.

### Story Length

Story length is determined by the overall number of nodes included in the story. In order to ensure a stable story viewer experience, we do not alter the length of the story transition based on the content. The story nodes and transitions have a consistent scroll length. As transitions at times contain complex camera movement and transitions, they are equally important as image and video presentation. Story pacing is solely determined by the scrolling speed of the story viewer. The overall story length can be influenced by the overall website height. We provide a pre-defined website height where we allocate 3000 pixels for each story node, including in- and out-transitions. This value can be changed on demand by the story author. As we also allow for dynamic narratives, the visible scroll bar extent loses importance in the story presentation as shorter paths in a story result in the scroll bar indicator not reaching the end of the scroll bar. To remedy this, we introduce a concept that replaces the scroll bar in our story presentation, similar to how the Genexpressionen story handles scrolling indication [29].

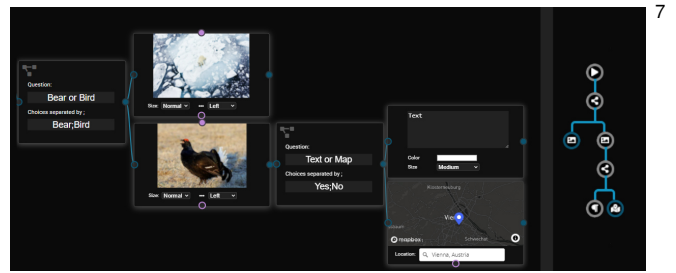

Fig. 5. An example story is visible on the left with the associated story tree visualization on the right.

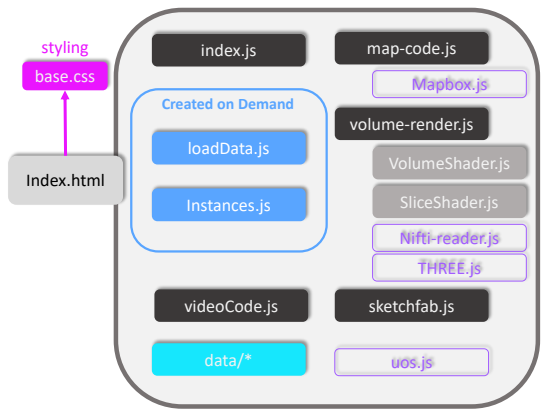

Fig. 4. Exported website structure: index.html is the starting point for the scrollytelling website. In blue, we present nodes where code is generated on the fly and in gray we present nodes which have been developed by us, but are story-independent. Violet nodes reflect imported external libraries.

### Dynamic Narratives

To enable exploration of dynamic narratives on the web with reversible decisions, we introduce a real-time on demand story node linkage method. In the exported web page, the story is created on the fly. Every story node is linked on demand to the next one and the story transitions are defined in real-time in order to enable dynamic narratives. If a decision node is encountered by a viewer, they can select a path and the selected branch is traversed. If the viewer scrolls back previous traversed story nodes are "unhooked" allowing for rapid story transitions in both directions in real-time. This flexible approach allows for reversible dynamic narratives on the fly while allowing for smooth and uninterrupted scrollytelling viewer experiences.

### Story Tree View

According to Kosara [22], it is important for scrollytelling viewers to get a sense of the overall story length and where they are currently positioned within the overall story. As visible in the example story in Figure 5 on the right-hand side, we present a story overview. The viewer can see how many segments the story consists of, if there are decision steps, and which media types are present. In addition, they can see what story point they are currently viewing, how many steps have been traversed, and how many there are still to explore. The story tree view consists of nodes representing story segments depicting node type as an icon, where the root node is visualized at the top and subsequent nodes are placed underneath predecessor nodes. Furthermore, decision nodes show up as splits into story branches. The viewer can always track what part of the story they are currently exploring, and to which point they have to scroll back to explore another decision path. In Figure 5, an example story and the accompanying story tree view are visible. While scrolling through the website, the outer ring of the story nodes the viewer has already visited will turn gray from the original blue color one by one. This keeps track of previous decisions and allows viewers to scroll back up to select other unexplored paths.

## 5 Implementation

In Figure 4, the overall structure of exported websites is presented. In *ScrollyVis*, we build up upon a variety of different JavaScript libraries. *Drawflow* [40] delivers the base functionality of dragging and dropping nodes onto a canvas for further linking up with other nodes. It provides standard interactions for placing and linking nodes. Furthermore, we customize the existing XML export functionality for the serialization of our stories. In order to fulfill our requirements, we added several features and altered specific behaviors of the library. First, we added previews to all nodes in order to enable storytellers to preview the final result of the website while editing. We also implemented support for drop-downs and check-boxes within nodes for setting specific parameters and the ability to customize connection ports in order to realize our layering approach. Another major addition is the context menu that is depicted in Figure 3 on the right. It can be customized for each node type and features only nodes which are compatible with the associated node port.

The volumetric visualization is based on WebGL and we leverage Three.js [5] as a basis for both direct volume rendering and slice views. We use custom vertex and fragment shaders to enable direct volume rendering in Three.js, based on code provided by Valentin Demeusy [11]. To allow for volume rendering in Three.js [5], we added shaders for volume and slice rendering to the library using WebGL. For the volume shader, we currently support maximum intensity projection, isosurface and direct volume rendering. When including map views, we use *Mapbox* and *OpenStreetMap*. We use JSZIP [21] for dynamic website packing making the website ready for downloading. For reading NIfTI files, we use NIFTI-READER-JS developed by Jack L. Lancaster and Michael J. Martinez [25], [26], [27]. We utilize uos.js to support scrollytelling in our approach, which is provided by Colin van Eenige [14]. To allow for efficient 3D surface visualization we integrate *Sketchfab* [39] by using the Sketchfab API. We plan to make our ScrollyVis editor freely available so that interested parties can generate their own scrollytelling web pages for scientific communication and outreach activities.

## 6 Case studies

We invited researchers from three different scientific fields to create stories about their work together with us. For further details, we refer to the additional materials for high resolution images of the networks created and video versions of the websites. Interactive additional materials, including the case studies as websites, are available at the following link: ScrollyVis use cases[1] . For our scenarios, we invited an osteology expert, a meteorology visualization expert and a PhD student in anatomy education. Two of the experts have a doctoral degree in their respective fields and several years of experience and one is a PhD student. For our case studies, we created a guided dynamic narrative about their research and gathered their impression of our prototype, the presentation-, and interaction-style. In addition, we invited them to share their views on the potential of scrollytelling as a tool for scientific communication, education, and outreach to a more general audience. As a first case study we present a recreation of a story by The Economist.

---

1. https://ericmoerthuib.github.io/ScrollyVis/

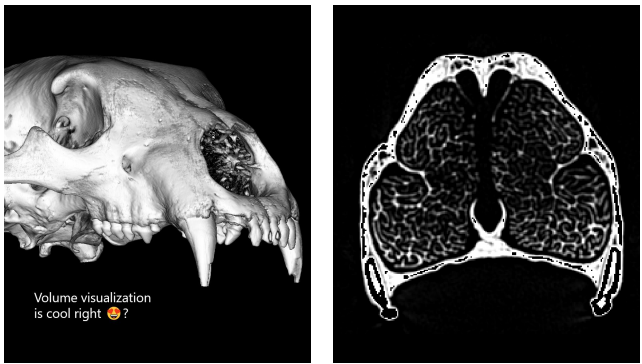

Fig. 6. Direct iso-surface volume rendering and slice volume rendering of scanned animal skulls in the osteology story co-created with an expert.

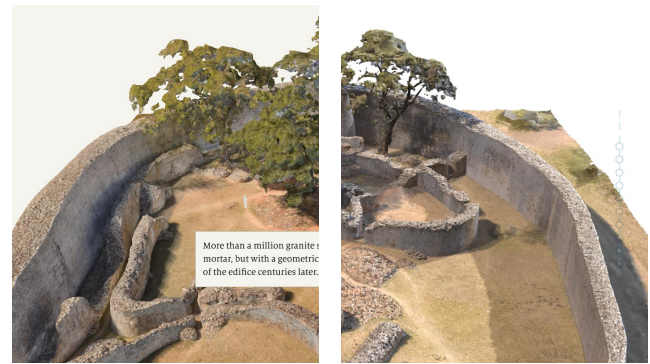

Fig. 7. An archaeological story by The Economist including 3D models of the Great Enclosure and the Hill Complex by the Zamani Project of the University of Cape Town. On the left a screenshot of the original story and on the right side our version.

## 6.1 The Economist: Unearthing the Truth

In order to exemplify that our approach is capable of reproducing professionally developed scrollytelling content, we use the story mentioned in section 3 by The Economist. It employs 3D models of the great enclosure and the hill complex, created by the Zamani Project of the University of Cape Town, South Africa. Creating such websites involves extensive web development skills and takes a substantial amount of time to create. To show the utility of our editor, one of our authors created a similar story with our editor in 15 minutes using models available on Sketchfab (see Figure 7). The story is included in the additional materials and available at the following link: Our Story[2], the original story is available under the following link: The Economist [13].

## 6.2 Scientific Outreach: Osteology Research

Together with the first expert, an Associate Professor at the University Museum of Bergen and the curator of the modern osteological collections, we have formulated a case study focused on outreach activities. Part of the story content is highlighted in Figure 6. The volumetric data used in this story were acquired by a Computed Tomography (CT) scanner. The story primarily focuses on polar bears in Norway, inspired by a blog post on this topic explaining characteristics of their skulls in relation to their habitat. In the story, we include several decision nodes to support viewer engagement and educational goals. Furthermore, we import custom geo-spatial data, consisting of polar bear sightings in Norway around Svalbard. The goal of the story is to showcase the potential of using our tool as a scientific outreach tool and for creating interactive web-based museum exhibits.

First, we introduced our ScrollyVis editor to the osteology expert and invited her to create a story completely on her own including various media types like a map view or a volume visualization. The expert has some experience in working with volume visualization and gave us the feedback that our interaction methods are as simple as the other tools she uses in her work. Furthermore, she created a volume animation consisting of several steps and told us that she has never created such an animation before. After this phase of the evaluation, we created a story together with the expert and encouraged her to speak openly about potential advantages and disadvantages of our approach. She thinks that the story created with ScrollyVis has a high potential to excite and engage visitors of the University Museum. According to her, the questions presented in the story can engage the viewers. During our discussion, she added that it would be beneficial to make the story mobile phone-friendly such that museum visitors can re-experience or share their experience within the museum with friends and family. She thinks enhancing the exhibition with on-demand mobile- or touchscreen-based scrollytelling content has high potential to enrich the museum attending experience. In comparison to specifically designed museum exhibits that are developed together with visualization researchers, e.g., *Living Liquid* by Ma et al. [30] or *Sea of Genes* by Dasu et al. [10], creating a story with ScrollyVis is feasible without collaborating with external researchers or paying for professional services. She reflected that our editor is nicely designed and looks user-friendly, though she expects that an initial learning phase together with us might be helpful. All in all, she is excited about the potential of ScrollyVis for scientific outreach activities and would like to further explore the opportunities in her blog posts and potentially for an upcoming exhibition in the University Museum.

## 6.3 Meteorological Visualization

The second story is authored together with a visualization expert. She is specialized in environmental visualization and her latest publication features geo-spatial and meteorological data visualization. First, she created a story including map views and decision nodes, which is essential for communication of her work, on her own. Finally, we jointly created a story presenting one of her papers [12]. The paper presents *Hornero*, a visual analytics tool for the detection and characterization of hazardous thunderstorms. Meteorological visualization is a very powerful tool to analyze the potential effects of hazardous weather phenomena. Presenting this information to a targeted audience or the general public is crucial to limit life-threatening risks. In addition to providing analytics tools for meteorologists and expert forecasters, it is equally important to present the results in an easy and comprehensible way to a more general audience. Selected story elements from the Hornero story are shown in Figure 8.

In general, this expert liked our ScrollyVis editor for authoring the story about her work, but she thought it would be interesting to actually show the Hornero-based interactive visualizations in the story. To this end, we included custom code provided by the authors of Hornero to integrate their custom geo-spatial visualizations. She thinks that ScrollyVis is an exciting, accessible and easy way to create scrollytelling stories for the web. She has experience in creating websites and thinks that our approach greatly simplifies the process of producing high-quality and easily accessible content. In a recent project, she has used Figma to create mock ups and she thinks that one advantage of our approach is also that the full

2. https://ericmoerthuib.github.io/ScrollyVis/UnearthingTheTruth

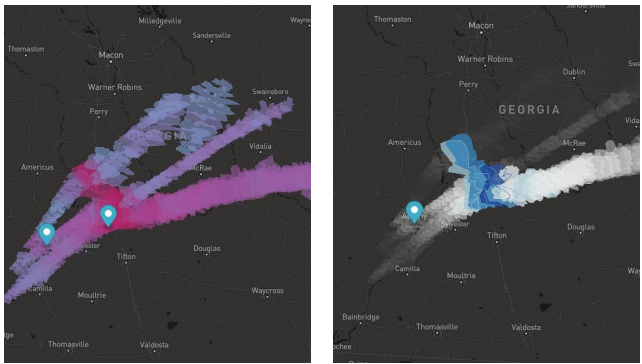

Fig. 8. A meteorological story including a storm strength and hail visualization using custom code provided by the paper authors.

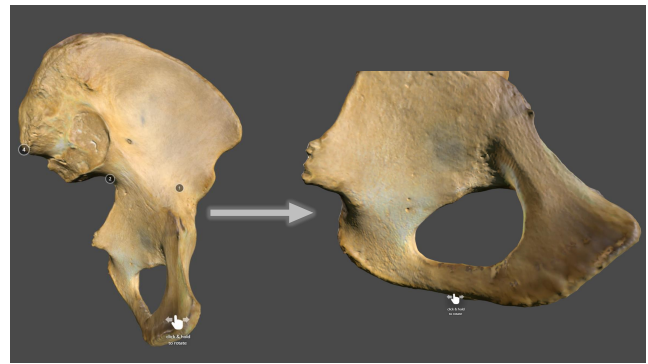

Fig. 9. ScrollyVis case exploring anatomy education potential based on specimen surface scans. The automatic camera transitions defined by starting and end point in the editor is particularly interesting in this case.

source code of the resulting website is available. In the future, she considers using our approach to explore the implications of interaction methods on communicating severe weather conditions to the general public. Similar to the first expert, she also requested improved support for mobile devices, as we currently mainly focus on desktop viewers. By simply altering the style sheet, we could already make the websites mobile friendly. Furthermore, she mentioned that our story tree view can be valuable for the story viewers but she would like to add keywords to the different nodes and would like them to be interactive and clickable. In conclusion, this expert thinks that ScrollyVis could be a part of many aspects of meteorological visualizations and can be used to communicate weather events in an exciting and engaging way.

### 6.4 Anatomy Education

The third expert we invited to use ScrollyVis is a PhD student who studies different means of anatomy education for medical students. She investigates different presentation media to analyze the effectiveness and learning outcomes. In their current setup, Sketchfab is used to show anatomical surface scans of bones which are made available within the online learning platforms the students are used to. The expert would like to use ScrollyVis to explore the effectiveness of guided dynamic narratives as teaching method compared to their existing content. She was invited to use ScrollyVis and was then asked to take part in an interview with us. The researcher was able to create stories on her own after watching the introduction video included in the additional material, without needing any support from us. She sees great potential in the editor and thinks the integration of Sketchfab enables them to create stories which might increase the learning outcomes of the medical students. Her advisor, a professor teaching anatomy and practicing orthopedic surgeon, appreciates the support for including audio files to add spoken commentary to interesting spots of the bone surface scans. This is a feature that is not available in the standard interface of Sketchfab. Furthermore, the expert would wish for a separate quiz mode where the students are not able to change answers after making a decision in a branching node.

## 7 EVALUATION

To evaluate the usability of our web-based authoring tool, we invited twelve people with diverse professional backgrounds. E1-E6 are masters students in computer science, E7-9 are PhD candidates in the field of visualization, E10 is a PhD candidate in the field of machine learning, and E11 holds a PhD degree in cell biology and is a post doc in the field of cancer imaging since 2018. Finally, E12 holds a PhD degree in medical imaging physics. None of the participants were involved in the development of our tool.

In the beginning of the evaluation, we invited the participants to watch a ten minute long ScrollyVis introduction and a tutorial video that is included in the additional materials. After that, participants were invited two create two different stories with the editor. The first story had the goal to familiarize the study participants with the different node types and interactions the editor provides by asking them to construct a story about a black grouse. In this story, the participants could still ask for help if they got stuck or had questions regarding the editor. The requirements for this story were that it has to include the following node types: decision, image, video, text and volume visualization. If these requirements were fulfilled, the first story was finished. The second story authoring task was more specific and included map views, image views, a decision node and text views. The participants also had to define both main paths as well as sub-paths to fulfill the task requirements. For this part of the evaluation, we only monitored the work of the participants but did not help them to reach their goal. Furthermore, we measured the time it took the attendants to fulfill the requirements. We used a think-aloud protocol where we asked participants to vocalize their thoughts and share their experience while they created the stories. After the study, the participants were invited to answer a questionnaire with 21 questions regarding different aspects of ScrollyVis. In addition to our evaluation form, the participants filled out the system usability scale (SUS) designed by Brook et al. [4]. All statements are evaluated based on a 5-point Likert scale with some statements negatively formulated.

### 7.1 Evaluation Results

The results of the evaluation are shown in Table 1. All questions which are marked with a star are negatively formulated and we present them here in their positive form with inverted responses for ease of interpretation. Overall, the study participants provided positive feedback about the editor. The lowest average value in our study is 3.92 out of 5 regarding the volume visualization methods and the two most positive responses with 5 out of 5 are about the integration of videos in a story and the usefulness of the context menu. The study participants were positive about the preview windows we integrated in the editor, but some wished for a more accurate depiction of how things will be aligned on the final website. The current functionality of placing text in horizontal and vertical alignment presets was sufficient for the participants, but some wished to see where the text box will be displayed in the node preview. One participant mentioned that he thinks some help was needed during the creation of the website, but overall, the score for that point is 4.17 out of 5. One evaluation participant suggested that some previous knowledge on working with storyboards is helpful. Another participant thought that the decision node usage could be improved by highlighting the options along the paths exiting the decision node. All participants generally agreed that they do not need prior web development knowledge (4.92/5) and Response of the participants on a 5-point Liker scale, where 1: strongly disagree, 2: disagree, 3: neither agree nor disagree, 4: agree and 5: strongly agree. Statements marked with a star are rephrased to the positive form in this table with inverted scores for presentation purposes. In the rightmost column, average values are presented. The second to last row reveals the results of the SUS questionnaire. The last row shows the time it took the participants to create the second evaluation story.

| | Statements | S1 | S2 | S3 | S4 | S5 | S6 | P1 | P2 | P3 | P4 | O1 | O2 | Avg. |
|---|---|---|---|---|---|---|---|---|---|---|---|---|---|---|
| G1 | I would like to use the ScrollyVis editor for creating Scrolltelling websites | 4 | 5 | 4 | 5 | 5 | 5 | 4 | 5 | 4 | 5 | 5 | 4 | 4.58 |
| G2 | Interacting with the editor is straightforward * | 5 | 4 | 4 | 5 | 5 | 4 | 5 | 4 | 5 | 4 | 5 | 4 | 4.50 |
| G3 | I don't need any web development pre-knowledge to create a website with the editor * | 5 | 5 | 4 | 5 | 5 | 5 | 5 | 5 | 5 | 5 | 5 | 5 | 4.92 |
| G4 | Creating a story based on the nodes and links in the editor is straightforward | 4 | 5 | 4 | 5 | 3 | 5 | 5 | 5 | 4 | 4 | 4 | 4 | 4.25 |
| G5 | I don't need help using the editor to create a Scrolltelling website in addition to the provided tutorial | 5 | 5 | 4 | 4 | 4 | 5 | 4 | 5 | 2 | 5 | 3 | 4 | 4.17 |
| N1 | The node preview (see image) helps me to find the right visualization settings for each file | 5 | 5 | 4 | 5 | 4 | 4 | 5 | 3 | 5 | 5 | 5 | | 4.50 |
| N2 | Adding a map view with a specified location was easy for me * | 5 | 5 | 5 | 5 | 5 | 5 | 5 | 5 | 3 | 5 | 5 | 4 | 4.75 |
| N3 | Using ScrollyVis, I don't need pre-knowledge about shaders and WebGL to create a volume visualization on the web | 4 | 5 | 3 | 5 | 5 | 4 | 5 | 5 | 3 | 5 | 5 | | 4.50 |
| N4 | I can easily integrate a video in my story | 5 | 5 | 5 | 5 | 5 | 5 | 5 | 5 | 5 | 5 | 5 | 5 | 5.00 |
| N5 | Combining different node types (e.g., map, text, image, . . . ) is easy * | 5 | 4 | 5 | 3 | 5 | 5 | 5 | 5 | 4 | 5 | 5 | 5 | 4.67 |
| I1 | The preview window in each node helps me to imagine the resulting website while editing it | 3 | 4 | 4 | 4 | 5 | 4 | 3 | 4 | 5 | 5 | 4 | 4 | 4.08 |
| I2 | The clone interaction (see image) helps me to create an animation of volume data | 5 | 5 | 5 | 5 | 4 | 4 | 4 | 5 | 5 | 4 | 4 | 5 | 4.58 |
| I3 | The main path and sub path feature helps me to create more complex stories | 5 | 5 | 5 | 5 | 5 | 4 | 4 | 5 | 3 | 4 | 3 | | 4.42 |
| I4 | The context menu (see image) helps me to create stories more efficiently * | 5 | 5 | 5 | 5 | 5 | 5 | 5 | 5 | 5 | 5 | 5 | 5 | 5.00 |
| I5 | Linking the nodes to create the story I would like to tell is easy and self-explanatory | 3 | 5 | 3 | 5 | 4 | 5 | 4 | 5 | | 5 | 5 | | 4.25 |
| I6 | Creating a story with the editor does not need pre-existing knowledge about designing storyboards * | 5 | 5 | 4 | 5 | 5 | 5 | 2 | 5 | 5 | 5 | 5 | | 4.58 |
| I7 | Adding questions for the viewer to decide which path the story shall go is easy and intuitive | 4 | 4 | 4 | 5 | 5 | 4 | 5 | | 2 | 5 | 5 | | 4.42 |
| I8 | The interaction methods used to define the visual exploration are intuitive and easy to use | 4 | 4 | 5 | 2 | 3 | 3 | 4 | 4 | 5 | 4 | 4 | | 3.92 |
| R1 | The tree view on the final website helps me to know in which path of the story I am. | 4 | 5 | 5 | 5 | 5 | 3 | 4 | 5 | 3 | 5 | 5 | | 4.33 |
| R2 | The resulting webpage reflects the intention I had when designing the story in the editor * | 5 | 5 | 5 | 5 | 5 | 5 | 4 | 5 | 4 | 5 | 5 | 5 | 4.75 |
| R3 | The viewer interaction (decisions) is nicely integrated on the website | 5 | 5 | 5 | 5 | 5 | 4 | 4 | 5 | 3 | 4 | | | 4.33 |
| SUS | System Usability Scale | 90 | 97,5 | 82,5 | 100 | 87,5 | 82,5 | 80 | 100 | 90 | 90 | 92,5 | 95 | 90,63 |
| T | Time to create the second story | 04:36 | 05:39 | 03:36 | 03:50 | 02:56 | 02:46 | 04:42 | 09:30 | 04:18 | 05:17 | 07:20 | 08:21 | 05:14 |

most participants would like to use the editor in the future (4.58/5). We further received positive feedback about the clone feature and that the resulting website reflects the intention of the story creators (4.75/5).

**System Usability Scale Scores**: Our SUS scores are presented in the last row of the evaluation result in Table 1. The results range from 82,5 to 100. On average, our application reached a SUS score of 90.63. Bangor et al. [1] introduced different ways of interpreting SUS scores, including the acceptability range, a grade scale, and an adjective rating scale. ScrollyVis achieved the highest score possible in all three categories: the acceptance rate is Acceptable, the grade scale score is A, and the application received an adjective rating of Excellent.

**Story Creation Time**: The user evaluation included a section where participants created a well-defined story on their own after familiarizing themselves with the editor. For this part of the evaluation, we report the average and individual story creation time. The average story creation time was 5 minutes and 14 seconds. The shortest time was 2 minutes and 46 seconds and the longest was 9 minutes and 30 seconds. In our results, the master students of computer science were the quickest, followed by the PhD students. The two participants with backgrounds in medical physics and cell biology took slightly longer to create their stories.

We conclude from this initial study that ScrollyVis was considered useful by students and researchers of various backgrounds. In general, ScrollyVis received positive feedback from all participants and the most of them would like to use the editor in the future for various tasks. For example, some stated they would like to create a review of their achievements in the last year or to present the latest research results. The participants reported that the editor features all necessary and relevant features to use it for creating scrollytelling websites. Still, the participants had feature requests to improve the user experience even more. However, none of the evaluation participants felt that features were currently missing in order to effectively work with the editor.

## 7.2 Expert Feedback

To gain further insights into the utility of our approach, we conducted an interview with an expert in scientific storytelling via interactive articles. He holds a PhD in Computer Science and Engineering and worked collaboratively with designers, developers, and scientists at Apple, Microsoft Research, and the NASA Jet Propulsion Lab. We presented the expert our editor and all capabilities it provides and afterwards we discussed the stories created in our case studies. In general, the expert found that current tools capable of creating stories comparable to our results are cumbersome and involve extensive knowledge of web development skills. One similar approach called *Idyll* [6] focuses mainly on parameter exploration and does not support as many media types out of the box as our editor does. According to the expert, the support of immersive media like 3D volumetric and surface data is unique and makes our editor stand out compared to other approaches. Furthermore, the non-linear path support makes the stories more engaging and interesting to explore. The expert thought that the stories we presented were of high quality and was impressed by the ease of use of our editor. Normally, developing such scrollytelling experiences involves different skills and may include teams of 6 to 12 people, whereas with our editor it is feasible for one person alone to create a similar story if suitable assets are already available. Furthermore, the expert thought that scrollytelling websites are an important aspect of scientific communication and he believes that our editor could make significant contributions to the developments in this field.

## 8 Discussion

Our approach can generate guided interactive narratives in a scrollytelling environment. In our prototype authoring tool, authors without programming experience can create such interactive narratives with simple drag, drop, and linking interactions. During our case studies with experts from different fields, we noted high engagement both in viewing and authoring stories. Our approach has potential for outreach to the general public and as well as for more specialized applications such as medical education. For example, scientific communication goals could be achieved by including ScrollyVis-authored stories on the web to advertise museum exhibitions or to communicate research results. Other potential use cases for ScrollyVis are the creation of additional materials for scientific papers and anatomy education based on scanned anatomical specimens. ScrollyVis could be a part of paper submission materials as well as 'science in plain English' presentations of scientific results. While scrollytelling in general might not be the best fit for all visualization goals, in particular for tasks that are of a more exploratory or analytical nature, we believe that in combination with guided dynamic narratives it can be an effective way to present information. As demonstrated by our use cases, all experts are excited about the potential ScrollyVis offers and have concrete plans for use of ScrollyVis in the future.

One potential avenue of improvement frequently mentioned by the experts is limited support for mobile devices in the current prototype implementation. One solution would be to use server-based rendering for some of the more computationally heavy visualization techniques. In addition, the style sheet for the exported websites is currently not designed for mobile devices. Furthermore, we would like to explore the possibilities of including a Latex and Python interpreter to our editor in able to allow researchers to present their mathematical formulas in the same way as in their papers. A Python interpreter would be beneficial as it would simplify the creation of information visualization on the fly in a form many researchers are used to. Currently, we support all dynamic narrative structures introduced by Munday [32], but some improvements could be made in the context of concentric narratives, where it may be convenient for viewers to automatically return to the initial decision node after having reached the end of a story segment.

## 9 CONCLUSION AND FUTURE WORK

With ScrollyVis, we introduce an authoring approach for realizing guided dynamic narratives as scrollytelling websites. We designed a extensible web-based story authoring tool that exports results ready for deployment on a web server. Our approach utilized a hybrid node-link storyboard editor which allows storytellers to get a good understanding of the resulting story during the authoring phase. Furthermore, a story tree view is available during story viewing that shows the story extent at a glance and where in the story viewers currently are. Finally, our story nodes support a variety of different media types including images, videos, audio, interactive maps, direct volume rendering and surface visualizations. Our system was designed with extensibility in mind, allowing for the easy integration of additional content.

We present a quantitative user evaluation with twelve independent participants with various professional backgrounds. Overall, ScrollyVis got a positive response from the study participants. The System Usability Scale was on average at the best possible grade level. The participants would like to use the editor in the future and think that all relevant features are included to effectively work with the editor. Furthermore, we present four case studies, three of them collaboratively authored with experts from three different scientific disciplines. Overall, all experts were highly engaged in both authoring and viewing the stories. They expressed interest in using our approach in future projects, ranging from blog posts to creating additional publication materials. Furthermore, we invited an expert in the creation of interactive online articles to qualitatively evaluate the ScrollyVis editor and the quality of the created stories. The expert thinks that our approach has great potential and fills a gap in the scrollytelling editor landscape. The quality of the stories matches those created by big news agencies, but the creation does not require a large team of web developers. In general, the expert is convinced that ScrollyVis is valuable and useful for scientific outreach.

With this work, we demonstrate the potential for our approach to create immersive guided dynamic scrollytelling web experiences without having to write any code. We are confident that our authoring tool is a basis for further research in guided dynamic narrative structures and their effectiveness for scientific communication and educational purposes. Viewer engagement is a top priority for storytellers and with ScrollyVis we empower authors to create dynamic stories and explore their effect on viewers. In the future, we would like to extend our authoring tool and provide default templates to further lower the barrier to getting started with ScrollyVis. These templates could provide general structures used in different story types and the authors can then bring in custom content. We also envision a collaborative editing process where multiple authors can work on the same story. Furthermore, we would like to enhance the tree view with interactive navigation actions to directly access various story elements and to add further node transition possibilities between nodes of different types. With our work, we aim to inspire people to share their stories as we are confident that everyone has a story worth telling.

## ACKNOWLEDGMENTS

This research was funded by the Trond Mohn Foundation (grant numbers '811255' and '813558').

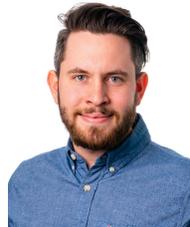

**Eric Mörth** is a PhD candidate in the visualization research group at the Univ. of Bergen, Norway since 2019. He is also affiliated with the Mohn Medical Imaging and Visualization (MMIV) center at the Haukeland University Hospital. In his PhD, he researches visualization techniques for multimodal medical imaging data. He received his masters' degree in Medical Informatics in 2018 from the Medical University of Vienna and his master's degree in Biomedical Engineering in 2019 from the Technical University of Vienna.

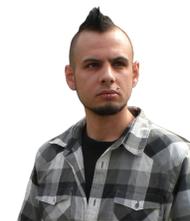

**Stefan Bruckner** is a full professor in Visualization at the Dept. of Informatics of the Univ. of Bergen, Norway. He received his master's degree (2004) and Ph.D. (2008), both in Computer Science, from the TU Wien, Austria, and was awarded the habilitation (venia docendi) in Practical Computer Science in 2012. Before his appointment in Bergen in 2013, he was an assistant professor at the Institute of Computer Graphics and Algorithms of the TU Wien. His research interests include all aspects of data visualization, with a particular focus on interactive techniques for the exploration and analysis of spatial data.

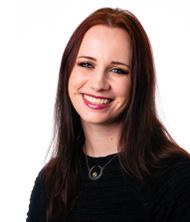

**Noeska N. Smit** is an associate professor in Medical Visualization at the Dept. of Informatics at the Univ. of Bergen, Norway, since 2017. She is also affiliated with the Mohn Medical Imaging and Visualization (MMIV) center as a senior researcher at the Haukeland University Hospital. After working as a radiographer for three years, she completed her studies in Computer Science in 20212 at the Delft University of Technology, the Netherlands, and in 2016, she obtained her PhD in medical visualization at the same institute. Currently, she researches novel interactive visualization approaches for multimodal medical imaging data.